\documentclass[useAMS,usenatbib]{mn2e}
\usepackage{graphicx}
\usepackage{amssymb}
\usepackage{url}

\usepackage{txfonts}
\usepackage{color}
\usepackage{natbib}
\usepackage{tabularx}

\title{A history of the gamma-ray burst flux at the Earth from Galactic globular clusters}

\author[W. Domainko et al.]
  {W.~Domainko,$^1$
  C.A.L.~Bailer-Jones,$^2$ F.~Feng,$^2$\\
  $^1$Max-Planck-Institut f\"ur Kernphysik, P.O. Box 103980, D-69029 Heidelberg, Germany\\
  $^2$Max-Planck-Institut f\"ur Astronomie, K\"onigstuhl 17, D-69117 Heidelberg, Germany}
\date{Xxxxx XX}

\pagerange{\pageref{firstpage}--\pageref{lastpage}} \pubyear{20xx}

\def\LaTeX{L\kern-.36em\raise.3ex\hbox{a}\kern-.15em
    T\kern-.1667em\lower.7ex\hbox{E}\kern-.125emX}

\begin{document}

\label{firstpage}

\maketitle

\begin{abstract}
Nearby gamma-ray bursts (GRBs) are likely to have represented a significant threat to life on the Earth.  Recent observations suggest that a significant source of such bursts is compact binary mergers in globular clusters.  This link between globular clusters and GRBs offers the possibility to find time intervals in the past with higher probabilities of a nearby burst, by tracing globular cluster orbits back in time.  Here we show that the expected flux from such bursts is not flat over the past 550\,Myr but rather exhibits three broad peaks, at 70, 180 and 340\,Myr ago. 
The main source for nearby GRBs for all three time intervals is the globular cluster 47\,Tuc, a consequence of its large mass and high stellar encounter rate, as well as the fact that it is one of the globular clusters which comes quite close to the Sun. Mass extinction events indeed coincide with all three time intervals found in this study, although a chance coincidence is quite likely.  Nevertheless, the identified time intervals can be used as a guide to search for specific signatures of GRBs in the geological record around these times.

\end{abstract}

\begin{keywords}
 globular clusters: general -- Gamma-ray burst: general -- Galaxy: kinematics and dynamics -- Astrobiology
\end{keywords}

\section{Introduction}

Globular clusters, densely packed groups of old stars, can efficiently produce close stellar binaries by dynamical interactions of their member stars. Examples for such dynamically formed binaries include low-mass X-ray binaries \citep[e.g.][]{clark1975,katz1975}, cataclysmic variables \citep{pooley2006} and milli-second pulsars \citep[msPSRs,][]{ransom2008,abdo2010}.  The most extreme binaries found in globular clusters consist of two neutron stars \citep{anderson1990}. Mergers of such binaries are believed to be the central engine of short gamma-ray bursts (GRBs) \citep{grindlay2006,dado2009,lee2010}, that produce brief, intense flashes of ionising radiation.
In contrast to short bursts, long bursts are believed to originate from the death of short-lived massive stars \citep[see][for a review on long and short bursts]{gehrels2009}. It has been argued that the rate of short GRBs in the local universe is dominated by the merger of neutron star binaries formed in globular clusters \citep{salvaterra2008,guetta2009}.  A link between globular clusters and short GRBs is further supported by the presence of a short GRB remnant candidate in the Galactic globular cluster Terzan~5 \citep{domainko2011a}, observed in the very-high energy gamma-ray \citep{abramowski2011,abramowski2013}, X-ray \citep{eger2010,eger2012} and radio wave band \citep{clapson2011}.  Additional evidence for the GRB - globular cluster connection comes from spatial offsets of short GRBs from their host galaxies \citep{berger2010,salvaterra2010,church2011} and the redshift distribution of such events \citep{salvaterra2008,guetta2009}.

Since globular clusters follow well-defined orbits around the Galaxy \citep{domainko2011b}, their coupling with GRBs allows us to examine the long-standing question of the past history of gamma-ray flux on the Earth.  (A similar approach for supernovae exploding in star clusters has been used in \citet{svensmark2012}). Numerous studies have shown that gamma rays from supernovae (SNe) or GRBs could, in principle, have had a significant impact on the Earth's atmosphere and biosphere, potentially even contributing to mass extinctions (see \citet{thorsett1995,scalo2002,melott2004,thomas2005} for the affect of GRBs in general, and \citet{dar1998,melott2011} for the affect of merger-induced bursts).  However, demonstrating that SNe or GRBs may in fact have played some role first requires identifying that sources could have come near enough to the Earth at some point. Some previous studies have attempted to make a connection between the solar motion relative to the Galactic plane or spiral arms, on the assumption that the gamma ray flux incident on the Earth is larger in these regions of enhanced massive star formation rate and/or increased stellar density (see \citet{cbj2009} for a review). However, a recent study 
shows that the flux from these sources as modulated by the plausible solar motion over the past 550\,Myr has a
poor correlation with the variation of the extinction rate on the Earth (Feng \& Bailer-Jones, submitted).

Indeed, it seems that astronomical phenomena alone are unlikely to be the dominant driver of biological evolution or the cause of all (or even most) mass extinctions. 
Nonetheless, if a GRB were to explode near to the Earth, its consequences could be catastrophic, and globular clusters are presumably a significant source of GRBs. 

The goal of this paper is to reconstruct the orbits of globular clusters relative to the Sun in order to calculate the GRB flux at the Earth as a function of time, and thereby to identify potential candidate clusters.
The data for this orbital reconstruction comes from the positions, distance, proper motion and radial velocity catalogues of globular clusters of \citep{dinescu1997,dinescu1999a,dinescu1999b,dinescu2003}, from which we obtain the current Galactic coordinates and space velocities. By sampling over the (often significant) uncertainties in the reconstructed orbits of the globular clusters and the Sun, we infer the expected GRB flux as a function of time. This allows us to identify the most probable intervals in the Earth's history of a significantly increased gamma ray flux, which may (or may not) be associated with times of higher extinction rate.

In section~\ref{method:orbits} we describe the orbital reconstruction method, and in section~\ref{method:summing} we explain how we derive from this the probability distribution over the past cluster--Sun separation and the expected gamma ray flux at the Earth. This takes into account the different GRB rates in the clusters, which is derived in section~\ref{method:weighting}. We give our results in section~\ref{sec:results} where we also identify some past extinction events. We conclude in section~\ref{sec:conclusions} with an outlook on how to further this work.

\section{Methods}

\subsection{Reconstructing Galactic orbits}\label{method:orbits}

We trace the orbits of the Sun and the globular clusters back in time by integrating the equations of motion through  the Galactic potential. In a purely gravitational system there is no dissipation of energy, so the dynamics are reversible. We adopt an analytic, three component, axisymmetric potential, $\Phi$, comprising the Galactic bulge, halo and disk
\begin{equation}
\Phi(R,z)=\Phi_b+\Phi_h+\Phi_d \ . 
\label{eqn:potential}
\end{equation}
The bulge and halo are represented with a Plummer distribution
\begin{equation}
\Phi_{b,h}=\frac{- G M_{b,h}}{\sqrt{R^2+z^2+b_{b,h}^2}}
\label{eqn:pot-bh}
\end{equation}
in which the characteristic length scales are $b_b=0.35$\,kpc for the bulge and $b_h=24.0$\,kpc for the halo,
and the bulge and halo masses are $M_b=1.40 \times 10^{10}$\,M$_\odot$ and
$M_h=6.98\times 10^{11}$\,M$_\odot$ respectively. 
$R$ is the radial coordinate perpendicular to the axis, and $z$ is the distance from the Galactic plane.
For the disk we use the potential from \cite{miyamoto75}
\begin{equation}
\Phi_d= \frac{- G M_d}{\sqrt{R^2 + \left(a_d+\sqrt{z^2+b_d^2}\right)^2}}
\label{eqn:pot-d}
\end{equation}
with the values $M_d=7.91 \times 10^{10}$\,M$_\odot$ for the disk mass, and $a_d=3.55$\,kpc and $b_d=0.25$\,kpc for the scale length and scale height of the disk, respectively (after \cite{sanchez01}).  The integration is performed numerically 
from the present back to 550\,Myr BP (before present). This time limit is chosen because it corresponds to the beginning of the Phanerozoic eon, a time from which the fossil record becomes more indicative of biodiversity variations.
The globular clusters (and Sun) are treated as massless.

\begin{figure}
\centering
\includegraphics[width=0.5\textwidth]{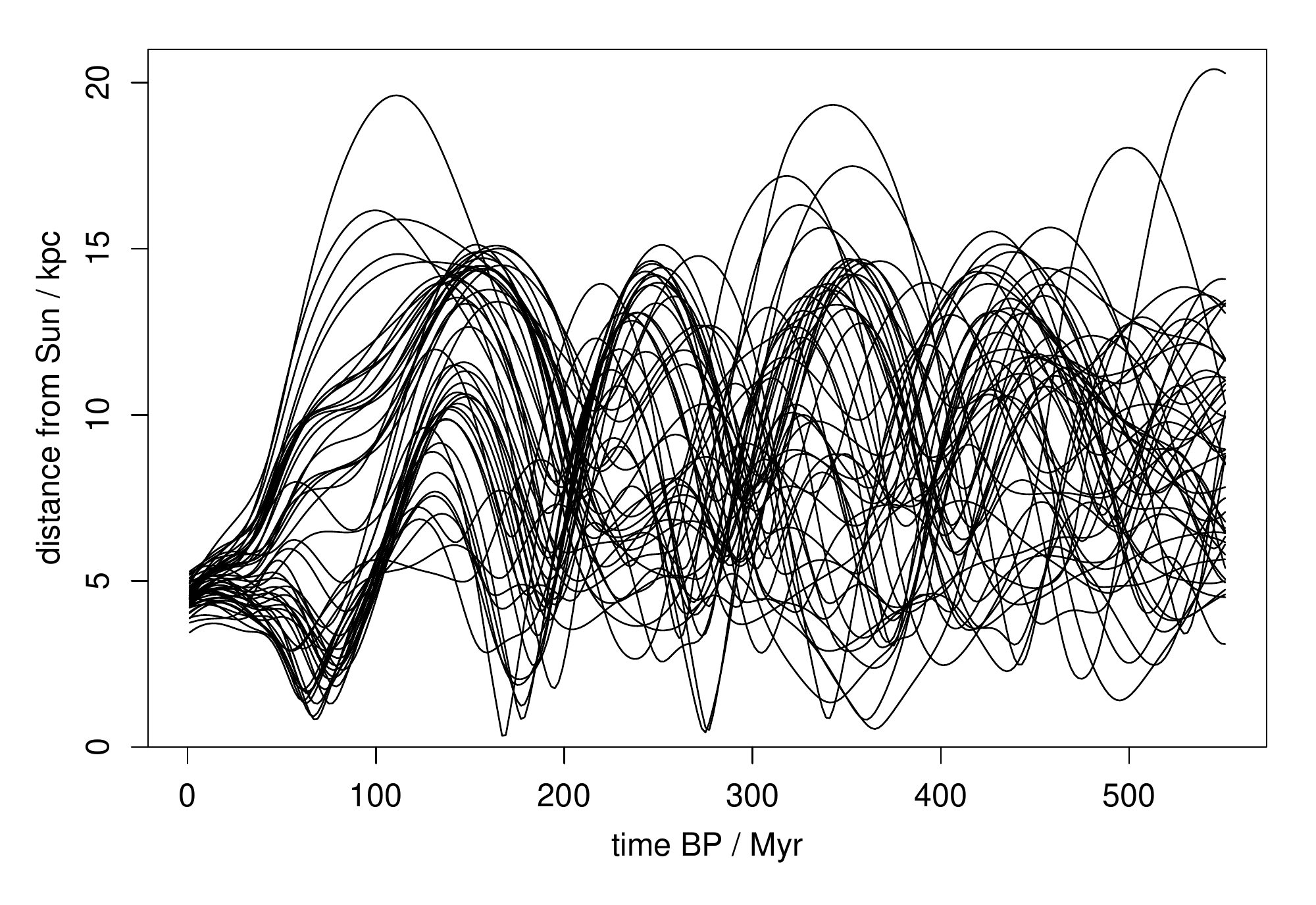}
\caption{Samples of the orbit of 47\,Tuc relative to the Sun to show how their separation varies over time.
The variance arises from sampling the uncertainty in the current phase space coordinates of both the globular cluster and the Sun, and integrating each back in time through the Galactic potential.
}
\label{fig:orbits_1} 
\end{figure}

The initial conditions for the integration are the current phase space coordinates (three position and three velocity components) of the globular clusters (and Sun). These of course have significant uncertainties, each represented as a Gaussian with known mean (measured coordinate) and standard deviation (estimated uncertainty).  These come from Dana Casetti-Dinescu's catalogue for globular cluster's three-dimensional space velocities (2012 version)\footnote{\url{http://www.astro.yale.edu/dana/gc.html}} for the globular clusters, and from Hipparcos data by 
\citep{dehnen98} for the Sun.
We further use a distance of the Sun to the Galactic center obtained from astrometric and spectroscopic observations of the stars near the 
supermassive black hole of the Galaxy \citep{eisenhauer2003} and the displacement of the Sun from the Galactic plane is calculated
from the photometric observations of classical Cepheids by \citet{majaess2009}. 
Rather than just performing a single integration for each object (cluster or Sun), we Monte Carlo sample its initial conditions from the uncertainty distribution in order to build up a large sample of orbits. Figure~\ref{fig:orbits_1} shows an example of such sample orbits for one globular cluster, 47\,Tuc, by plotting the distance of the cluster from the Sun over time. (We sample over the possible orbits of the Sun too.)
We do not take into account the (possibly significant) uncertainties in the Galactic potential. In principle we could adopt an uncertainty model for these parameters and marginalize over them also. But we choose to omit this in this first investigation.

Finally we have to note that compact binaries may be ejected from their parent cluster before they merge and produce a GRB \citep[e.g.][]{phinney1991,ivanova2008}. 
This effect will smear out the distribution of compact binaries around the producing cluster. The typical escape velocities for massive globular clusters are about 
50\,km\,s$^{-1}$, which is comparable to the present uncertainties of the globular cluster velocity.
Although over time the orbit of the ejected binary could deviate considerably from its parent cluster,
the uncertainty in its orbit is comparable to the uncertainty for its parent cluster, which we take into account.
We therefore choose to omit the issue of ejected GRB progenitors for this first investigation.
Furthermore, more massive clusters are better able to retain their binaries, and these are the clusters that preferentially produce GRBs (see Sec. \ref{method:weighting}).

\subsection{The probability distribution over globular cluster distances and the expected GRB flux at the Earth}
\label{method:summing}

\begin{figure}
\centering
\includegraphics[width=0.5\textwidth]{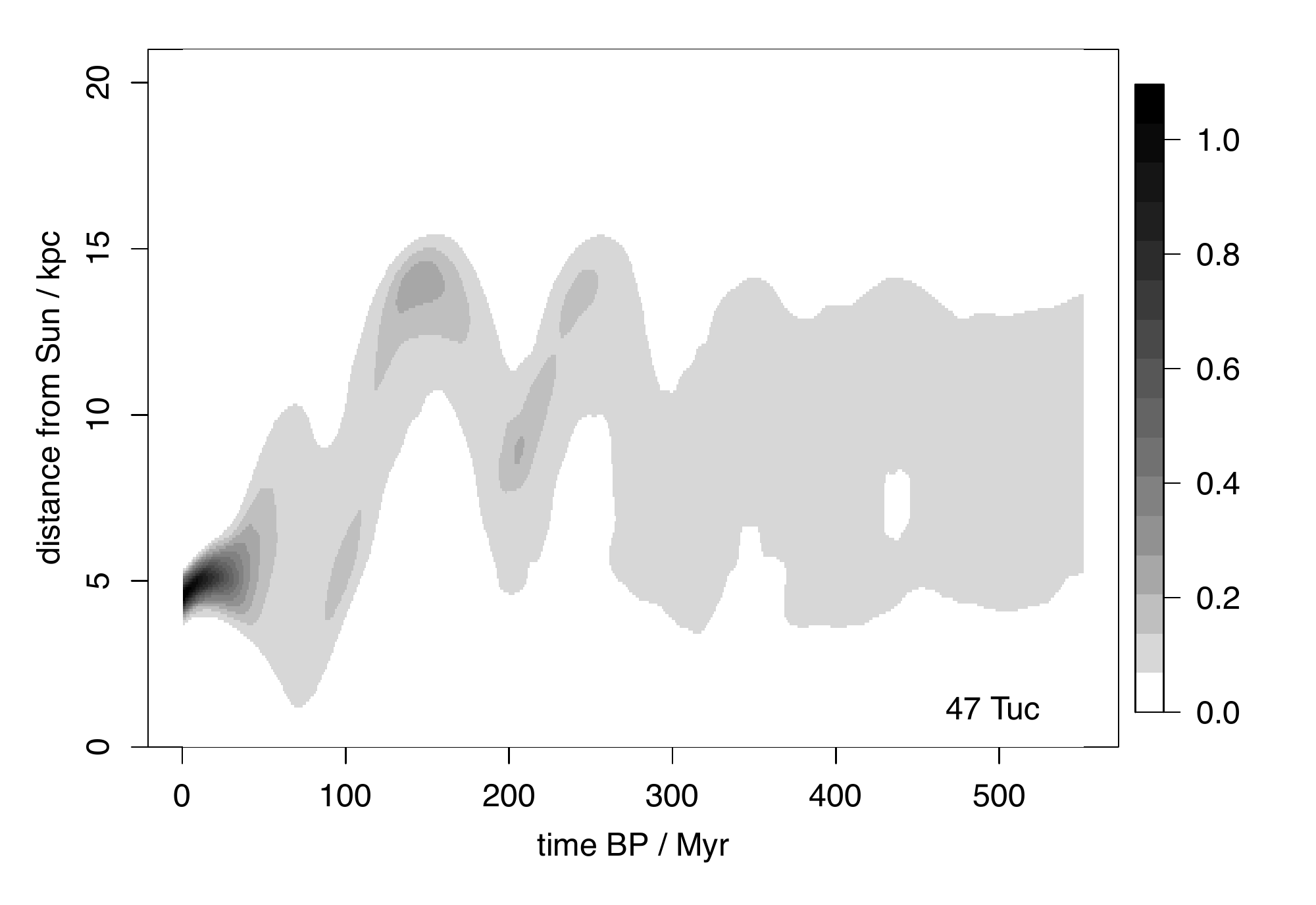}
\caption{The variation of the probability density, $f_c(r,t)$, of the distance $r$ between 47\,Tuc and the Sun as a function of time $t$, shown as a grey scale. This scale is normalized such that the integration over $r$ at each $t$ is unity.}
\label{fig:fcrt_1} 
\end{figure}

\begin{figure}
\centering
\includegraphics[width=0.45\textwidth]{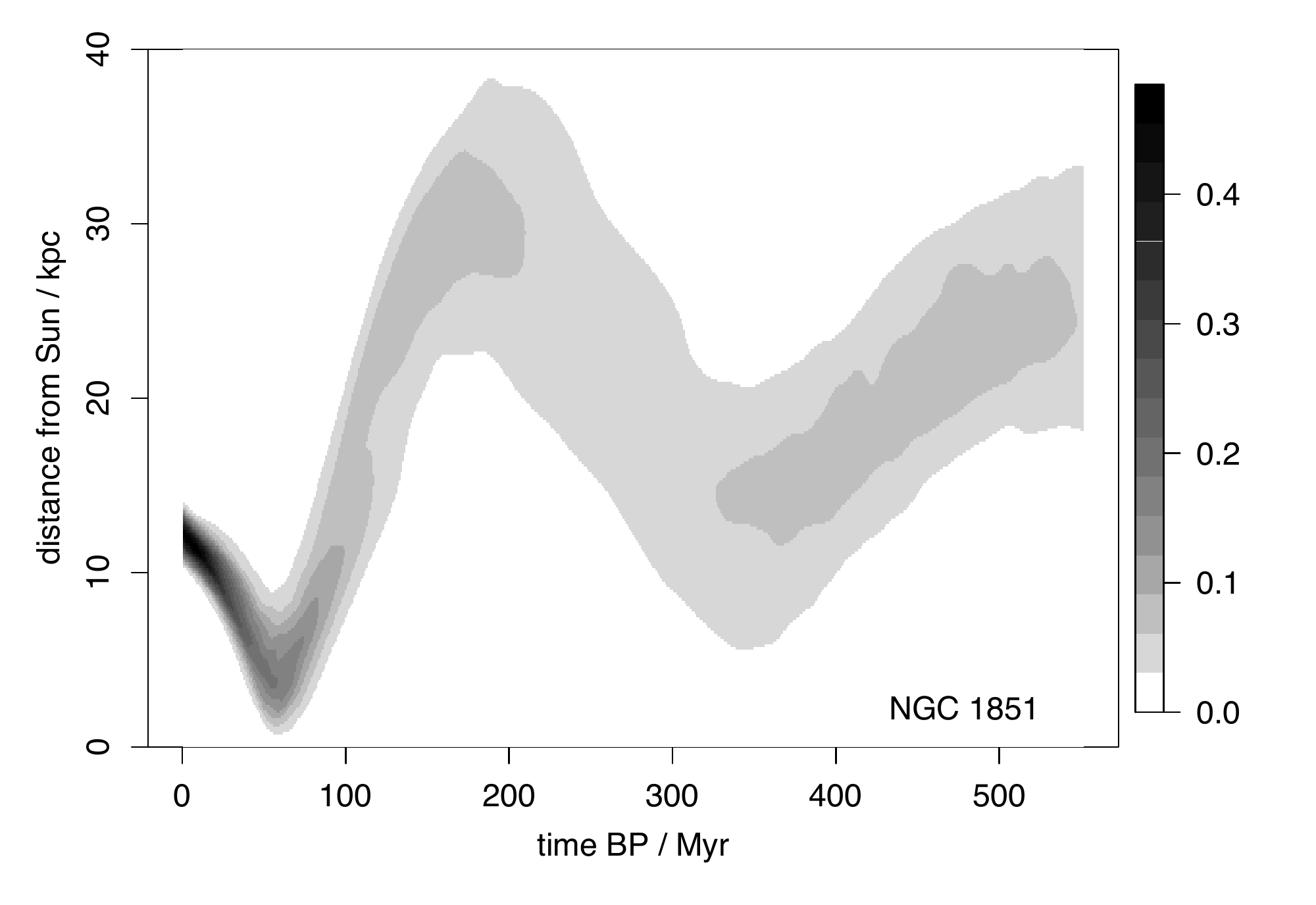}
\caption{As Figure~\ref{fig:fcrt_1} but for NGC\,1851}
\label{fig:fcrt_4} 
\end{figure}

\begin{figure}
\centering
\includegraphics[width=0.45\textwidth]{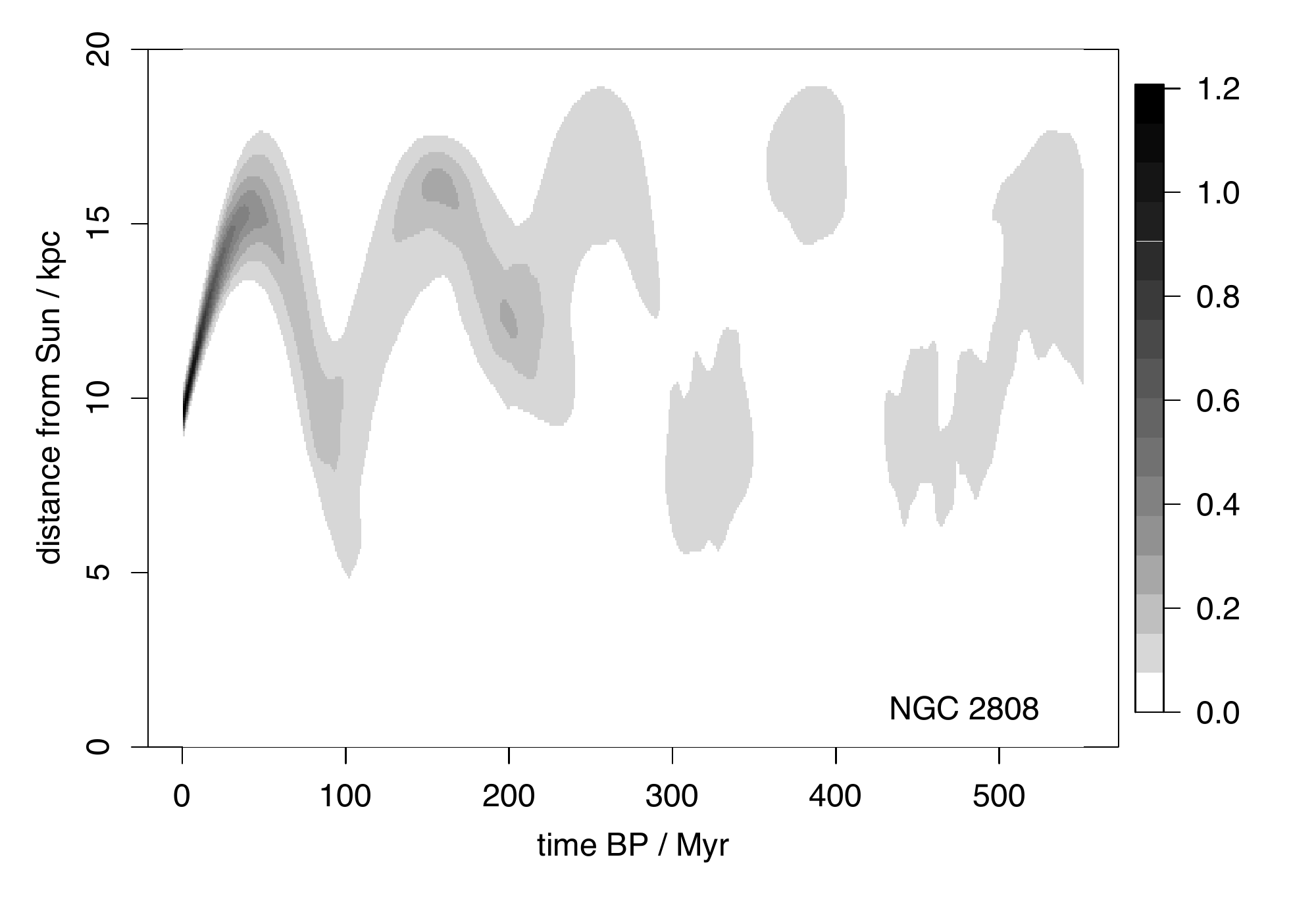}
\caption{As Figure~\ref{fig:fcrt_1} but for NGC\,2808}
\label{fig:fcrt_7} 
\end{figure}

\begin{figure}
\centering
\includegraphics[width=0.45\textwidth]{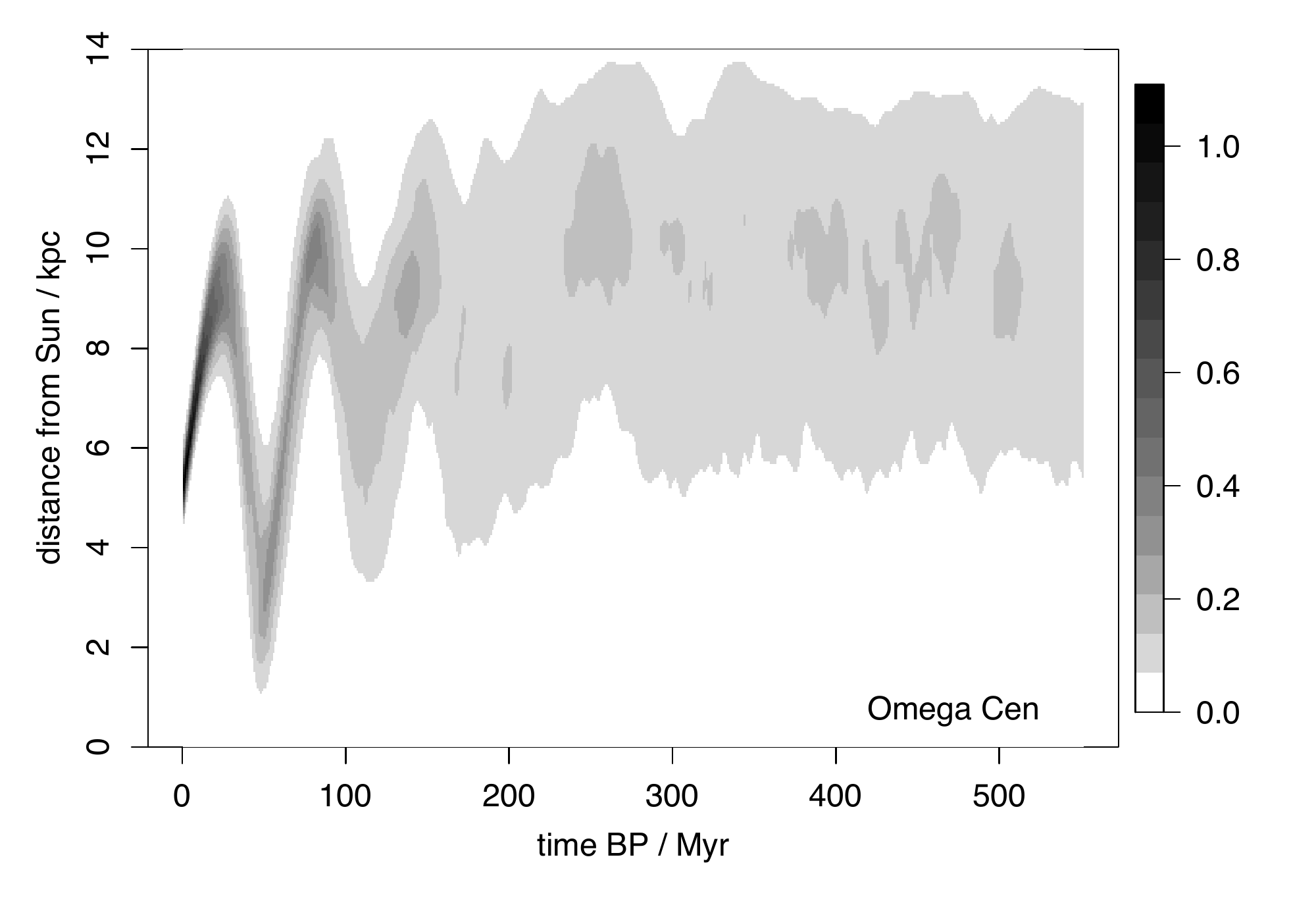}
\caption{As Figure~\ref{fig:fcrt_1} but for Omega Cen}
\label{fig:fcrt_14} 
\end{figure}

\begin{figure}
\centering
\includegraphics[width=0.45\textwidth]{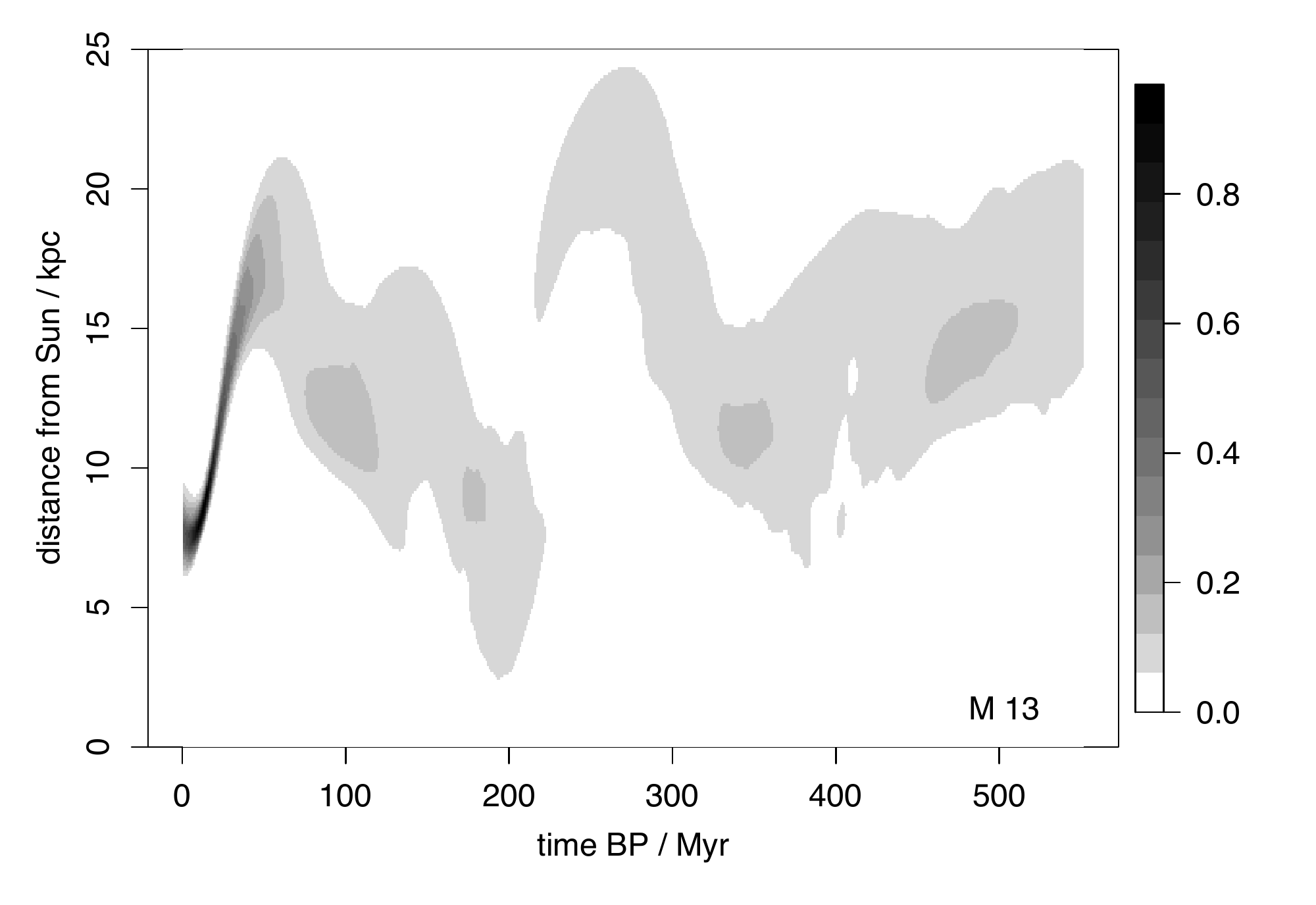}
\caption{As Figure~\ref{fig:fcrt_1} but for M\,13}
\label{fig:fcrt_26} 
\end{figure}

\begin{figure}
\centering
\includegraphics[width=0.45\textwidth]{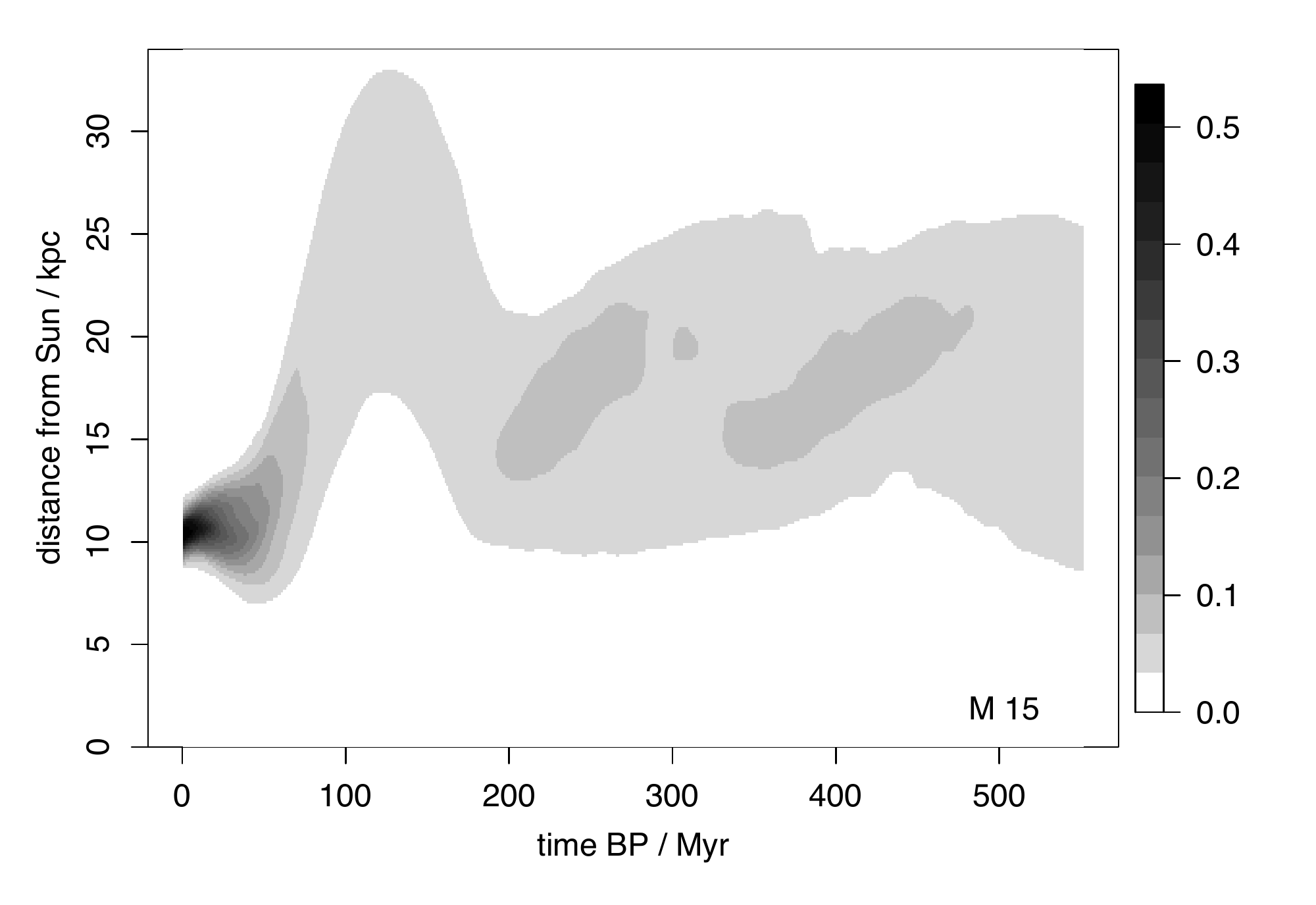}
\caption{As Figure~\ref{fig:fcrt_1} but for M\,15}
\label{fig:fcrt_55} 
\end{figure}

For a given globular cluster, $c$, we convert the set of (thousands of) relative orbits into a two-dimensional density distribution over time, $t$, and separation, $r$, using kernel density estimation.  We interpret the resulting distribution as a probability distribution of the Sun--cluster separation over time, $f_c(r,t)$, which is normalized such that $\int_r f_c(r,t) dr = 1$ for all $t$ and for each cluster.  This is shown in Figure~\ref{fig:fcrt_1} for 47\,Tuc, in which the probability density is plotted as a grey scale. At any given time, the darker the band, the more concentrated the probability is around a smaller range of distances.  The width of the distribution at any time is determined by how the uncertainties in the present coordinates of both globular cluster and Sun propagate back in time. The density estimates for some other globular clusters are shown in Figures~\ref{fig:fcrt_4}--\ref{fig:fcrt_55}.

The flux of a gamma ray burst at the Sun is proportional to $1/r^2$. Multiplying $f_c(r,t)$ by $1/r^2$, and assuming that gamma ray bursts occur at random times\footnote{GRBs are of course discrete, rare events. Lacking information on when they occurred, the best we can do is to derive the probability per unit time of a burst for each globular cluster.}, we get a 2D distribution which is proportional to the expected GRB flux from distance $r$ at time $t$. If we integrate this (at a time $t$) over all distances then we get a quantity, $\int_r \frac{1}{r^2} f_c(r,t) dr$, which is proportional to the expected GRB flux from that globular (at time $t$).  The important thing about this quantity is that it takes into account the uncertainties in the reconstructed globular cluster and solar orbits. 

We now extend this concept to the complete set of globular clusters. Each cluster has a different probability per unit time of producing a GRB, proportional to the factor $w_c$, defined in section~\ref{method:weighting}. We can then see that the quantity
\begin{equation}
\Psi(t) = \int_{r=0}^{r=r_\mathrm{max}} \sum_c w_c \frac{1}{r^2} f_c(r,t) dr
\end{equation}
is proportional to the expected GRB flux at the Sun at time $t$ from any globular cluster.  In principle we integrate up to $r_\mathrm{max}=\infty$, but in practice we can truncate it to a few kpc.  Indeed, if there is a minimum flux threshold below which the gamma ray flux is too small to have any significant affect on the Earth's biosphere or climate, then truncation is appropriate. 
Note that the absolute scale of $\Psi(t)$ is not calibrated: only relative values are meaningful.

\subsection{Weighting individual globular clusters}\label{method:weighting}

Observationally, the frequency of occurrence of GRBs in individual globular cluster is not known.  
The dynamical formation of compact binaries, proposed progenitors of such events, is rather complex,
involving at least two stellar encounters \citep[see][]{ivanova2008,ivanova2010}.
However, the rate of GRBs in each globular cluster is expected to be linked to the cluster properties. 
Several authors have already investigated the dependence of the compact binary formation rate on the characteristics of the clusters. 
\citet{ivanova2008} found that the formation of close double neutron star binaries depends on the square of the cluster density, and that the number of retained 
neutron stars increases as the escape velocity (and thus cluster mass) increases.
\citet{grindlay2006} used a model where the formation of double neutron star binaries scales linearly with the neutron star number density, the velocity dispersion (and thus mass of the cluster) 
and the number of potential progenitor systems (binaries containing one neutron star). Both models find that massive clusters with a high concentration of stars strongly favour the formation
of prospective GRB progenitor systems. Here we adopt a similar approach to these previous works and scale the expected GRB rate with quantities that are known for a large sample of globular clusters.

Specifically, assuming that GRBs are caused by neutron star encounters, then the GRB rate will depend on the number of neutron stars in the cluster and their encounter rate.  We assume that the number of neutron stars scales linearly with the mass of the globular cluster, $m_\mathrm{c}$, and thus linearly also with the cluster luminosity.  The total encounter rate, $\Gamma_c$, is given as $\Gamma_c \propto \rho_0^{1.5} r_\mathrm{core}^2$ \citep{pooley2006}, where $\rho_0$ is the central stellar number density and $r_\mathrm{core}$ is the core radius of the globular cluster. Values for these parameters for our sample of clusters we obtained from 
\citet[][2010 edition]{harris1996}\footnote{\url{http://physwww.mcmaster.ca/~harris/mwgc.dat}}.  Combining these two factors we get a quantity $w_\mathrm{c} = m_\mathrm{c} \Gamma_c$, which is proportional to the frequency of gamma rays bursts in the clusters, and is used as the weighting factor in section~\ref{method:summing}.  Accordingly, and as already noted in the beginning of this section, massive clusters with high concentrations of stars at their center have a large GRB rate.
We investigated the uncertainties of our approach by applying an alternative weighting scheme for individual globular clusters. We followed \citet{ivanova2008} and adopted
weights proportional to $\rho_0^2\, m_\mathrm{c}$. With this approach we found that the typical uncertainties for the leading clusters is a factor of a few, with a few notable exceptions 
(see Sec. \ref{sec:results}). For the results in Sec. \ref{sec:results} we use the weights $w_\mathrm{c}$ as defined earlier in this section.

Having calculated the indivdual weights, $w_\mathrm{c}$, they are then normalised such that the sum of all weights equals 1. Here we used 141 clusters from \citet[][2010 edition]{harris1996}
where all necessary parameters are known.
This, in principle, further allows us to estimate the expected absolute GRB rates for individual globular clusters by defining that a weight of 1 corresponds to the Galactic rate of GRBs launched in globular clusters. This galactic GRB rate can be calculated from the short GRB rate in the local Universe of $8_{-3}^{+5}$\,Gpc$^{-3}$yr$^{-1}$ \citep{coward2012} and the density of Milky Way-type galaxies of 0.01 Mpc$^{-3}$ \citep{cole2001}. 
This rate is obtained for GRBs beamed towards Earth and is thus independent of the degree of collimation of the events.
If it is assumed that the occurrence of short GRBs in the local Universe is dominated by bursts launched in globular clusters \citep{salvaterra2008,guetta2009}, then the combined GRB rate of all globular clusters is 10$^{-6}$~year$^{-1}$. This estimate is also consistent with the theoretically expected rate of short GRB production in these clusters \citep{lee2010}.

\section{Results}\label{sec:results}
 
\begin{figure}
\centering
\includegraphics[width=0.5\textwidth]{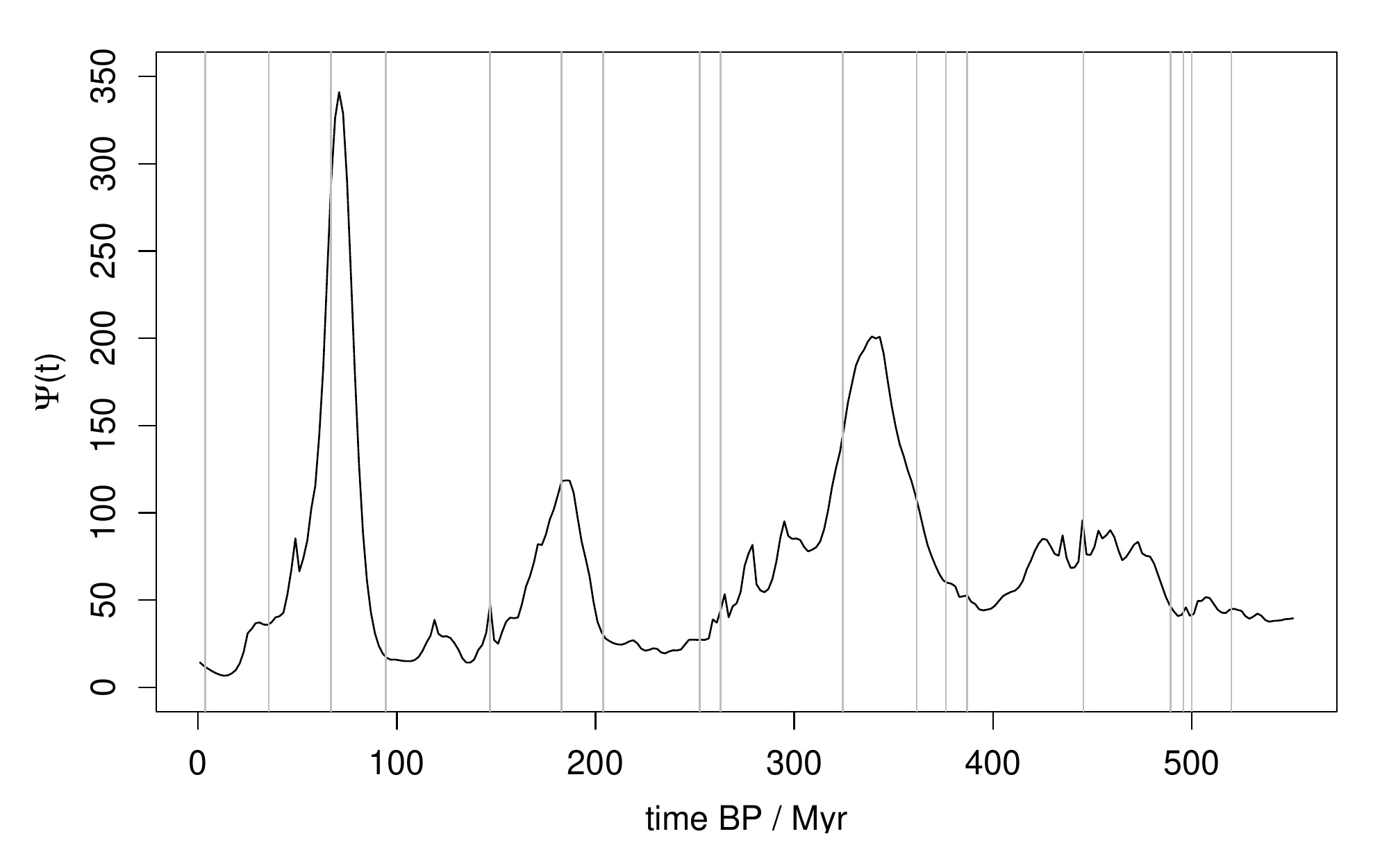}
\caption{The expected GRB flux, $\Psi(t)$, at the Sun as a function of time before present, in arbitrary units. The vertical lines are the times of the 18 mass extinction events compiled by \citep{bambach2006}. }
\label{fig:dwcipd} 
\end{figure}

Figure~\ref{fig:dwcipd} shows the expected GRB flux, $\Psi(t)$, for the case $r_\mathrm{max}=5$\,kpc.  This distance threshold covers 95\% of all hazardous GRBs if a log-normal GRB luminosity distribution with $\mathrm{log}\, E_\mathrm{\gamma,iso} = 50.81 \pm 0.74$~erg \citep{racusin2011}
and a critical fluence at Earth for a significant affect on the biosphere or climate
of $10^7$~erg~cm$^{-2}$ \citep{melott2011} is assumed.  
(The profile of $\Psi(t)$ has very similar shape for other values of $r_\mathrm{max}$, the difference being that 
the ``background'' level is higher for larger values of $r_\mathrm{max}$, and lower for smaller values.)  We see a significant variation.  There are three broad peaks, 
at 70, 180 and 340\,Myr.  These correspond to times in the Earth's history when -- within the limitations of our orbital reconstruction and assumptions made -- we would 
expect a significantly higher level of GRB flux than the average over the past 550\,Myr.

Examining the plots of $f_c(r,t)$ for all clusters, we can identify those clusters which make the biggest contribution to $\Psi(t)$ in each peak:

\begin{itemize}

\item{Peak at 70\,Myr}. 
The main contributor is 47\,Tuc, which has ten times the contribution to $\Psi(t)$ than does the next cluster, NGC\,1851

\item{Peak at 180\,Myr}.
The main contributor is again 47\,Tuc, with several others contributing at a level 5--20 times lower, the largest of these being Omega\,Cen, M\,13, and M\,15.

\item{Peak at 340\,Myr}.
Once again 47\,Tuc gives the largest contribution, with several others contributing at a level 7 or more times lower, the most significant of these being NGC\,2808.

\end{itemize}
The prominence of 47\,Tuc is a consequence both of its high weight, $w_c$, and the fact that it is one of the globular clusters which comes quite close to the Sun. 
All the main contributors are massive clusters that contain significant populations of dynamically formed stellar binaries. Specifically:
\begin{itemize}
\item{47\,Tuc} has the second largest number of radio-detected msPSRs \citep[23,][]{ransom2008}, detected by \emph{Fermi-LAT} in high energy gamma-rays \citep{abdo2010}.
In our weighting scheme (see Sec. \ref{method:weighting}) it would account for about 5\% of the GRBs produced in globular clusters.
In the alternative weighting scheme (see Sec. \ref{method:weighting}) it  accounts for about 1\% of GRBs in globular clusters (for the following clusters this number 
is given in brackets). 47\,Tuc is the dominant globular cluster in our study for both weighting schemes.
\item{NGC\,1851} contains a msPSR in a very eccentric binary system with massive secondary \citep{freire2004}. This could account for about 2\% (1\%) of GRBs from globular clusters.
\item{NGC\,2808} is a massive globular cluster with complex evolutionary history \citep{piotto2007}. This could account for about 5\% (0.3\%) of GRBs from globular clusters.
\item{Omega\,Cen} is the most massive globular cluster in the Galaxy, detected by \emph{Fermi-LAT} \citep{abdo2010}. This could account for about 2\% (10$^{-3}$\%)
of GRBs from globular clusters. For this globular cluster the two different weighting schemes give the largest difference since it is a very massive cluster with a shallow density profile.
\item{M\,13} contains five radio-detected msPSRs \citep{ransom2008}. This could account for about 0.2\% (10$^{-3}$\%) of GRBs from globular clusters.
\item{M\,15} has a double neutron-star binary that will merge within a Hubble time \citep{anderson1990}, eight radio-detected msPSRs \citep{ransom2008}. 
This could account for about 6\% (2\%) of GRBs from globular clusters.
\end{itemize}

As mentioned earlier, GRBs are of course discrete, rare events. Indeed, 
our calculations suggest that only about 10 GRBs will have occurred 
within 5kpc of the Sun over the course of the Phanerozoic. Thus the {\em 
true} distribution of flux with time would comprise of a series of 
narrow peaks of various heights. Fig. 8 shows the {\em expected} flux at 
time (times a constant), so is the best single estimate of that 
distribution.

By way of comparison we overplot in Figure~\ref{fig:dwcipd} the times of 18 mass extinction events on the Earth revealed by the fossil record, as compiled by \citep{bambach2006}. One may be tempted to draw a causal connection between one of these events and one of the peaks in $\Psi(t)$, although clearly there is a reasonable chance that one of these 18 events could coincide with a peak just by chance.\footnote{The probability that any one event, thrown down at random over the time range range 0--550\,Myr, would not land in a particular box of width 20\,Myr (the width of the peaks in $\Psi(t)$), is $1 - 20/550$. So the probability that at least one of the 18 events coincides with such a box just by chance is $1 - (1 - 20/550)^{18} = 0.5$. This is not a proper hypothesis test, but it highlights that a coincidence is quite likely.} It is nonetheless worthwhile identifying those events nearest to the three peaks. These are 
\begin{itemize}
\item{Peak at 70\,Myr}: the famous KT extinction at 65\,Myr\,BP, generally accepted to have had a significant role in the demise of the dinosaurs; 
\item{Peak at 180\,Myr}: the late Pliensbachian/early Toarcian (early Jurrasic) extinction event at 179--186\,Myr\,BP;
\item{Peak at 340\,Myr}: the early Serpukhovian (mid Carboniferous) extinction event at 322--326\,Myr\,BP,
and the late Famennian (late Devonian) extinction event at 359--364 Myr BP.
\end{itemize}
Whether or not a globular cluster GRB is implicated in any of these extinctions remains a subject for future work.

\section{Outlook}\label{sec:conclusions}

In this paper we have traced globular cluster orbits back to the beginning of the Phanerozoic eon in order to identify time intervals where a high flux of ionizing radiation caused by a nearby GRB is more likely.  We found that the probability for such an event is far from flat with time during the Earth's history. It instead exhibits several distinct peaks, the most prominent ones being around 70, 180 and 340\,Myr BP.  The main source of GRBs in all cases is 47\,Tuc.  All three time intervals can in principle be associated with a mass extinction event, although a chance coincidence is likely.  Therefore, to establish a link between a nearby GRB and an impact on the Earth and its biota, supporting geological signatures are needed.  Geological signatures could comprise radiation damage of crystals (e.g. fossil cosmic ray tracks \citep{fleischer1967} or color shifts \citep{ashbuugh1988}), deposition of radioactive isotopes \citep{dar1998} or elevated rates of bone cancer \citep{rothschild2003}.  The time intervals identified in this paper can be used as a guide-line to search for such signatures in the geological record.

Finally, the current orbital parameters of globular clusters and the solar system are subject to considerable uncertainties. 
(These were taken into account in our analysis, and contribute to smearing out the probability curve.)
This situation will be substantially improved in the near future with the launch of the Gaia satellite, which will determine the 
dynamics of the Galaxy with unprecedented accuracy. 
With better determined orbital parameters we will be able to constrain the past orbits more tightly, and so repeat this study to give results of higher confidence.

\label{lastpage}

\end{document}